\documentclass[amsmath,showpacs,aps,prx,reprint,floatfix]{revtex4-1}
\usepackage{graphicx, color, xfrac}
\usepackage{amsthm, amsmath, amssymb, bm, enumerate}
\usepackage[dvips]{epsfig}
\usepackage[dvips]{epsfig}
\usepackage[normalem]{ulem}

\newcommand{\pit}{{\tilde{\pi}}}

\begin{document}

\title{Momentum-Space Entanglement in Heisenberg Spin-Half Ladders}
\author{Rex Lundgren}
\affiliation{Department of Physics, The University of Texas at Austin, Austin, TX 78712, USA}

\date{\today}
\pacs{71.10.Pm, 03.67.Mn, 11.25.Hf}


\begin{abstract}
We analytically study momentum-space entanglement in quantum spin-half ladders consisting of two coupled critical $XXZ$ spin-half chains using field theoretical methods. When the system is gapped, the momentum-space entanglement Hamiltonian is described by a chiral conformal field theory with a central charge of two. This is in contrast to entanglement Hamiltonians of various real-space partitions of gapped-spin ladders that have a central charge of one. When the system is gapless, we interestingly find that the entanglement Hamiltonian consist of one gapless mode linear in subsystem momentum and one mode with a flat dispersion relation. We also find that the momentum-space entanglement entropy obeys a volume law.
\end{abstract}
\maketitle
\section{Introduction}

Quantum entanglement has become an indispensable tool in the study of condensed matter physics. In particular, the topological entanglement entropy \cite{PhysRevLett.96.110405,PhysRevLett.96.110404} and the entanglement spectrum \cite{Li:prl08} have played a significant part in understanding and identifying exotic phases of matter. The entanglement spectrum is obtained as follows: First a system is partitioned into two regions, $A$ and $B$. This partition is usually made in real-space. Given the reduced density matrix of $A$, $\rho_A$, where $\rho_A$ is obtained from the density matrix, $\rho$, (formed from the ground state wave-function) as $\rho_A=e^{-H_e}=\mathrm{Tr}_A(\rho)$, the entanglement spectrum is the set of eigenvalues of $H_e$, which is called the entanglement Hamiltonian. The entanglement entropy, $S$,  can be obtained from the entanglement spectrum as $S=\mathrm{Tr}\rho_A\mathrm{ln}\rho_A$. In a real-space partition, the entanglement entropy for topologically ordered states is equal to $\alpha |\partial A| -\gamma$, where $\alpha$ is a non-universal constant term, $|\partial A|$ is the length of the boundary between regions $A$ and $B$ and $\gamma$ is the topological entanglement entropy. The topological entanglement entropy has been studied in many interesting systems including quantum spin liquids \cite{PhysRevA.71.022315,PhysRevB.75.214407} and fractional quantum Hall systems \cite{Zozulya:prb07,PhysRevLett.98.060401,Laeuchli:njp10}. Most studies on the entanglement spectrum have used a bipartite real-space partition, including work on the entanglement spectrum of quantum spin chains \cite{Franchini:2010kq,2012JSMTE..08..011A,PhysRevLett.108.227201,Calabrese:pra08,Pollmann:prb10,Pollmann:njp10,2013arXiv1303.0741L,PhysRevB.88.125142,PhysRevB.90.235111,1742-5468-2014-6-P06001,1742-5468-2014-10-P10029} and ladders \cite{Poilblanc:prl10,2011EL.....9650006P,Lauchli:prb12,2012JSMTE..11..021S,Tanaka:pra12,Lundgren,PhysRevB.88.245137,Fradkin_Ladder}, fractional quantum Hall systems \cite{Li:prl08,Thomale_AC:prl10,Lauchli:prl10,PhysRevB.85.045119,Sterdyniak:prb12,Dubail:prb12,PhysRevLett.108.256806,PhysRevB.88.155307,PhysRevB.84.205136}, Chern insulators \cite{PhysRevX.1.021014,1742-5468-2014-10-P10030}, symmetry broken phases \cite{2013PhRvL.110z0403A,PhysRevB.88.144426}, topological insulators \cite{PhysRevB.82.241102,PhysRevLett.104.130502,PhysRevB.84.195103,Kargarian:prb10,PhysRevB.87.035119}, and other systems in one \cite{Deng:prb11,Turner:prb11} and two dimensions \cite{PhysRevLett.105.080501,Santos:prb13,PhysRevLett.107.157001,2013NJPh...15e3017S,1742-5468-2014-9-P09011,2014arXiv1410.4790S,2014arXiv1411.6932H}.

Recently, several works on the entanglement entropy and spectrum in one dimension have used a momentum-space partition. This partition is motivated in part by the low-energy description of one-dimensional systems, which involves splitting particles into right and left movers \cite{Giamarchi:book,Gogolin:book}. The momentum space entanglement spectrum was first studied in the bosonic formulation of the Heisenberg spin-half chain \cite{PhysRevLett.105.116805}, where it was found to reveal information about the underlying conformal field theory by the counting of entanglement levels and an entanglement gap. Ref.~\cite{2014arXiv1404.7545L} generalizes this work to the $XXZ$ model and studied the momentum-space entanglement spectrum of both fermionic and bosonic formulations of the $XXZ$ spin-half chain. For the bosonic formulation, Ref.~\cite{2014arXiv1404.7545L} finds that the entanglement gap seen in Ref.~\cite{PhysRevLett.105.116805} does not extend throughout the critical region of the $XXZ$ spin-half chain. For the fermionic formulation, the momentum-space entanglement Hamiltonian does not capture physical phase transitions. While this might seem like a drawback, Ref.~\cite{2014arXiv1404.7545L} highlights that these results might be useful for numerical algorithms, such as the momentum-space density matrix renormalization group. The momentum-space entanglement spectrum has also proved useful in characterizing disordered fermionic systems \cite{PhysRevLett.110.046806,ES_DIS}. Entanglement entropy between left and right movers was also recently studied in the context of string theory \cite{2014arXiv1407.7057P}. We note that the entanglement entropy of a momentum-space partition where fast modes were traced over, instead of partitioning left and right movers as done in this paper, was studied in Ref.~\cite{PhysRevD.86.045014}.

In this paper, we analytically study the momentum-space entanglement spectrum and entropy between left and right movers in Heisenberg spin-half ladders. The legs of the ladder consist of spin-half $XXZ$ chains of length $L$, described by the following Hamiltonian (with perodic boundary conditions):
\begin{equation}
H_{\alpha}=\sum_{i=1}^L \frac{J^{xy}}{2}(S^{+}_{\alpha,i}S^{-}_{\alpha,i+1}+S^{-}_{\alpha,i}S^{+}_{\alpha,i+1})+J^z S^z_{\alpha,i}S^z_{\alpha,i+1}
\end{equation}
where the leg index $\alpha=1,2$ and $J^{xy}>0$. We take the interchain coupling between the legs of the ladder to be
\begin{equation}
H_{\perp}=\sum_{i=1}^L\left( \frac{J^{xy}_{\perp}}{2}(S^{+}_{1,i}S^{-}_{2,i}+S^{-}_{1,i}S^{+}_{2,i})+J_{\perp}^zS^z_{1,i}S^z_{2,i}\right).
\end{equation}
$H_{\perp}$ couples spins that are on the same rung of the ladder. The total Hamiltonian is then $H=H_1+H_2+H_{\perp}$. By mapping the spin-ladder to a low-energy bosonic field theory and then expanding interchain interactions to quadratic order in fields, we are able to analytically obtain the momentum-space entanglement spectrum and entropy between left and right movers. If the system is gapped, we notably find the entanglement spectrum is gapless and has a central charge of two. This is in contrast to the entanglement Hamiltonian with a central charge of one that has been found in various real-space partitions of gapped spin ladders. These partitions include tracing out one leg of the ladder \cite{Poilblanc:prl10,2011EL.....9650006P,Lauchli:prb12,PhysRevB.88.245137,Fradkin_Ladder} and every other rung \cite{2014arXiv1408.1716S}. If the system is gapless, the entanglement Hamiltonian has one gapless mode and one dispersion-less mode. We find that the entanglement entropy scales with the length of the ladder, i.e. a volume law. This is also in contrast to the standard area law, which is usually seen in real-space systems \cite{RevModPhys.82.277}. We note that the entanglement entropy between coupled spin chains also scales with the length of the ladder \cite{PhysRevB.83.085112,COUPLEDCFT}.

Our paper is organized as follows: In section II, we introduce the low-energy field theory of the spin ladder under study. In section III, the momentum-space entanglement spectrum and entropy is calculated for this model. Finally, in section IV, we summarize our results and present our conclusion.

\section{Model}
We now present the low-energy model of the system. To begin, we first use the Jordan-Wigner transformation to map the spins to fermions. After this transformation, we express the fermionic operators in terms of bosons, i.e. we bosonize the fermionic system. See Ref.~\cite{Giamarchi:book} or Ref.~\cite{Gogolin:book} for a review of Abelian bosonization. The transformation is summed up in the spin-to-boson transformation
\begin{equation}
S_{\alpha}^+(x)=\frac{S_{\alpha,i}^+}{\sqrt{a}}=\frac{e^{i\theta_{\alpha}(x)}}{\sqrt{2\pi a}} ((-1)^x+\cos(2\phi_{\alpha}(x)))
\end{equation}
and
\begin{equation}
S_{\alpha}^z(x)=\frac{S_{\alpha,i}^z}{a}=-\frac{1}{\pi}\partial_x\phi_{\alpha}(x)+\frac{(-1)^x}{\pi a}\cos(2\phi_{\alpha}(x)).
\end{equation}

For this work, we consider the range of parameters for the legs of the ladder, $0\le\Delta=\frac{J^z}{J^{xy}}\le 1$. For this range of $\Delta$, the $XXZ$ spin-half chain is critical and the bosonized Hamiltonian of a single chain takes the form (after rescaling the fields by $\phi=\frac{\phi}{\sqrt{\pi}}$ and $\theta=\frac{\theta}{\sqrt{\pi}}$)
\begin{equation}
H_{\alpha}=\frac{u}{2}\int\mathrm{d}x \left(K(\partial_x\theta_{\alpha})^2+\frac{1}{K}(\partial_x\phi_{\alpha})^2\right)
\end{equation}
where
\begin{equation}
K=\frac{\pi}{2(\pi-\cos^{-1}(\Delta))},~u=J^{xy}\frac{\pi\sqrt{1-\Delta^2}}{2\cos^{-1}(\Delta)}.
\end{equation}
Each leg of the ladder is composed of left and right moving particles. At $K=1$, there is no entanglement between left and right movers \cite{Giamarchi:book,Gogolin:book}.

Introducing symmetric and anti-symmetric fields as follows
\begin{equation}
\phi_{\pm}=\frac{1}{\sqrt{2}}(\phi_1\pm\phi_2),~\theta_{\pm}=\frac{1}{\sqrt{2}}(\theta_1\pm\theta_2),
\label{SYM_ANTISYM_DEF}
\end{equation}
the total Hamiltonian is
\begin{align}
H=\frac{u_+}{2}\int\mathrm{d}x \left(K_+(\partial_x\theta_+)^2+\frac{1}{K_+}(\partial_x\phi_+)^2\right)\nonumber \\
+\frac{u_-}{2}\int\mathrm{d}x \left(K_-(\partial_x\theta_-)^2+\frac{1}{K_-}(\partial_x\phi_-)^2\right)+\nonumber \\
\frac{\pi J^{xy}_{\perp}a}{(2\pi a)^2}\int \mathrm{d}x \cos(\sqrt{2\pi}\theta_-)+\frac{J^z_{\perp}a}{(2\pi a)^2}\int \mathrm{d}x \cos(\sqrt{8\pi}\phi_-)+\nonumber \\
\frac{J^z_{\perp}a}{(2\pi a)^2}\int \mathrm{d}x \cos(\sqrt{8\pi}\phi_+)+\nonumber \\
\frac{1}{2}\frac{\pi J^{xy}_{\perp}a}{(2\pi a)^2}\int \mathrm{d}x \cos(\sqrt{2\pi}\theta_-)\cos(\sqrt{8\pi}\phi_+)
\label{H_TOTAL}
\end{align}
where
\begin{equation}
K_\pm=K\left(1\pm\frac{K J^z_{\perp} a}{\pi u}\right)^{-\frac{1}{2}},~u_\pm=u\left(1\pm\frac{K J^z_{\perp} a}{\pi u}\right)^{\frac{1}{2}}.
\end{equation}
In this work, we assume that $|J^{xy}_{\perp}|,|J^{z}_{\perp}|\ll J^{xy}$. For a wide range of $\Delta$, this model describes the Haldane phase for ferromagnetic rung coupling and the rung-singlet phase for anti-ferromagnetic rung coupling \cite{Giamarchi:book,Gogolin:book,PhysRevB.60.15230}. There are three cases of interest based on the relative scaling dimensions of the cosine terms. The scaling dimensions of $\cos(\sqrt{2\pi}\theta_-), ~\cos(\sqrt{2\pi}\theta_-),~\cos(\sqrt{8\pi}\phi_+),$ are $(2K_-)^{-1},~2K_-,~2K_+$ respectively. We consider each of the cases separately.
\subsection{$J^{xy}_{\perp}\neq0,~J_\perp^z\neq0,~K_-\ge\frac{1}{2}$}\label{sec:A}
We first consider the case when $J^z_\perp\neq0$. For $J_\perp^z\neq0$, the last term in Eq.~\eqref{H_TOTAL} is less relevant in the renormalization group sense than the other interaction terms and we ignore it. The symmetric and anti-symmetric modes are now separate. For $K_-\ge\frac{1}{2}$, the most relevant operator is the $\cos(\sqrt{2\pi}\theta_-)$ term \cite{Giamarchi:book}, thus we drop the $\cos(\sqrt{8\pi}\phi_-)$ term (Strictly speaking, at $K_-=\frac{1}{2}$, both cosine terms are equally relevant, but it is a useful approximation to neglect the $\cos(\sqrt{8\pi}\phi_-)$ term due to SU(2) symmetry present at $K_-=\frac{1}{2}$ \cite{PhysRevB.86.094417}.). The effective Hamiltonian for this range of parameters is then $H^A=H_+^A+H_-^A$, where
\begin{align}
H^A_+=\frac{u_+}{2}\int\mathrm{d}x \left(K_+(\partial_x\theta_+)^2+\frac{1}{K_+}(\partial_x\phi_+)^2\right)\nonumber \\
\frac{J^z_{\perp}a}{(2\pi a)^2}\int \mathrm{d}x \cos(\sqrt{8\pi}\phi_+)
\end{align}
and
\begin{align}
H^A_-=\frac{u_-}{2}\int\mathrm{d}x \left(K_-(\partial_x\theta_-)^2+\frac{1}{K_-}(\partial_x\phi_-)^2\right)+\nonumber \\
\frac{\pi J^{xy}_{\perp}a}{(2\pi a)^2}\int \mathrm{d}x \cos(\sqrt{2\pi}\theta_-)
\end{align}
For $\Delta$ between $0$ and $1$, both the symmetric and anti-symmetric channels are energetically gapped.
\subsection{$J_\perp^z\neq0,~K_-<\frac{1}{2}$}\label{sec:B}
We now consider the case when $K<\frac{1}{2}$. This set of parameters also includes the case when $J^{xy}_\perp=0$. The most relevant interaction term in the anti-symmetric channel is now the $\cos(\sqrt{8\pi}\phi_-)$ term \cite{Giamarchi:book}, and we ignore the $\cos(\sqrt{2\pi}\theta_+)$ term. We note that the symmetric channel is always gapped for the range of $\Delta$ considered. However, the anti-symmetric channel is not gapped for the whole range of $\Delta$ considered. For small $J^{z}_{\perp}$ (and $J^{xy}_{\perp}=0$), $\cos(\sqrt{8\pi}\phi_-)$ is irrelevant for around $K\approx1$ and thus the anti-symmetric channel is gapless \cite{PhysRevB.60.15230}. More concretely, if $K<1-\frac{J_{\perp}^z}{\pi u_-}$ the anti-symmetric channel is gapped. The effective Hamiltonian for this range of parameters is then $H^B=H_{+}^B+H_-^B$, where
\begin{align}
H^B_+=\frac{u_+}{2}\int\mathrm{d}x \left(K_+(\partial_x\theta_+)^2+\frac{1}{K_+}(\partial_x\phi_+)^2\right)\nonumber \\
\frac{J^z_{\perp}a}{(2\pi a)^2}\int \mathrm{d}x \cos(\sqrt{8\pi}\phi_+)
\end{align}
and
\begin{align}
H^B_-=\frac{u_-}{2}\int\mathrm{d}x \left(K_-(\partial_x\theta_-)^2+\frac{1}{K_-}(\partial_x\phi_-)^2\right)+\nonumber \\
+\frac{J^z_{\perp}a}{(2\pi a)^2}\int \mathrm{d}x \cos(\sqrt{8\pi}\phi_-).
\end{align}

\subsection{$J_\perp^z=0$}\label{sec:C}
We finally consider the case when $J^z_{\perp}=0$, which requires a bit more care as less relevant operators become important. The Hamiltonian now reads,
\begin{align}
H=\frac{u_+}{2}\int\mathrm{d}x \left(K_+(\partial_x\theta_+)^2+\frac{1}{K_+}(\partial_x\phi_+)^2\right)\nonumber \\
+\frac{u_-}{2}\int\mathrm{d}x \left(K_-(\partial_x\theta_+)^2+\frac{1}{K_-}(\partial_x\phi_+)^2\right)+\nonumber \\
\frac{\pi J_{\perp}^{xy}a}{(2\pi a)^2}\int \mathrm{d}x \cos(\sqrt{2\pi}\theta_-)+\nonumber \\
\frac{1}{2}\frac{\pi J_{\perp}^{xy}a}{(2\pi a)^2}\int \mathrm{d}x \cos(\sqrt{2\pi}\theta_-)\cos(\sqrt{8\pi}\phi_+).
\end{align}
Under renormalization group flow, the $\cos(\sqrt{2\pi}\theta_-)$ term will increase, while the $\cos(\sqrt{2\pi}\theta_-)\cos(\sqrt{8\pi}\phi_+)$ term initially decreases \cite{PhysRevB.60.15230}. As pointed out in Ref.~\cite{PhysRevB.60.15230}, this means $\theta_-$ essentially becomes pinned and we can write the following effective Hamiltonian for the symmetric channel for this range of parameters as
\begin{align}
H^C_+=\frac{u_+}{2}\int\mathrm{d}x \left(K_+(\partial_x\theta_+)^2+\frac{1}{K_+}(\partial_x\phi_+)^2\right)\nonumber \\
+\frac{\langle\cos(\sqrt{2\pi}\theta_-)\rangle}{2}\frac{\pi J_{\perp}^{xy}a}{(2\pi a)^2}\int \mathrm{d}x \cos(\sqrt{8\pi}\phi_+).
\end{align}
The expectation value of $\lambda=\langle\cos(\sqrt{2\pi}\theta_-)\rangle$ is taken with respect to
\begin{align}
H^C_-=\frac{u_-}{2}\int\mathrm{d}x \left(K_-(\partial_x\theta_+)^2+\frac{1}{K_-}(\partial_x\phi_+)^2\right)+\nonumber \\
\frac{\pi J^{xy}_{\perp}a}{(2\pi a)^2}\int \mathrm{d}x \cos(\sqrt{2\pi}\theta_-)
\end{align}
We thus see the effective Hamiltonians in Sec.~\ref{sec:A} and~\ref{sec:C} are the same upon swapping $J_\perp^z$ with $\lambda \pi J_\perp^{xy}/2$ in the symmetric channel. As such, we only need to calculate the entanglement spectrum for the Hamiltonian in Sec.~\ref{sec:A} and~\ref{sec:B}.

\section{Momentum-Space Entanglement Spectrum and Entropy}
In this section, we calculate entanglement entropy and spectrum between left and right movers. These left and right movers are free only when the legs of the ladder are uncoupled and the legs of the ladder are at the $XX$ point ($\Delta=0$). This is equivalent to the momentum-space partition considered for the fermionic representation of the $XXZ$ chain in Ref.~\cite{2014arXiv1404.7545L}. We first note that the reduced density matrix of left or right movers can be factored as $\rho_A=\rho_{A}^{\mathrm{zero}}\otimes\rho_{A}^{\mathrm{osc}}$. Furthermore, we can factorize $\rho_{A}^{\mathrm{osc}}$ as $\rho_{A,\pm}^{\mathrm{osc}}\otimes\rho_{A,\pm}^{\mathrm{osc}}$. 

We now calculate the entanglement Hamiltonian for the Hamiltonians in Sec.~\ref{sec:A} and~\ref{sec:B}. We first introduce left and right moving fields, 
\begin{equation}
\phi_{\alpha}=\frac{\phi_{R,\alpha}+\phi_{L,\alpha}}{\sqrt{4\pi}},~\theta=\frac{\phi_{\alpha,L}-\phi_{\alpha,R}}{\sqrt{4\pi}}.
\end{equation}
When $K=1$, the left and right movers of a single chain are free. When $K\neq1$ there is finite entanglement between left and right movers, even if the legs of the ladder are uncoupled, in this basis. The left and right moving fields have the following mode expansion \cite{PhysRevB.88.245137}
\begin{align}
\phi_{\alpha,R}=\phi_{\alpha,R,0}+2\pi N_{\alpha,R}\frac{x}{L}+\nonumber \\
\sum_{k>0}\sqrt{\frac{2\pi }{L|k|}}\left(a_{k,\alpha}^{\dagger}e^{ikx}+a^{\phantom{\dagger}}_{k,\alpha} e^{-ikx})\right)
\label{ModeRight}
\end{align}
and
\begin{align}
\phi_{\alpha,L}=\phi_{\alpha,L,0}+2\pi N_{\alpha,L}\frac{x}{L}+\nonumber \\
\sum_{k<0}\sqrt{\frac{2\pi }{L|k|}}\left(a_{k,\alpha}^{\dagger}e^{ikx}+a^{\phantom{\dagger}}_{k,\alpha} e^{-ikx})\right).
\label{ModeLeft}
\end{align}
Here, $a_k$ are bosonic operators describing oscillator modes and $\phi_{\alpha,L/R,0}$ and $N_{\alpha,L/R}$  are the zero modes. The zero modes, $\phi_{\alpha,L/R,0}$ and $N_{\alpha,L,R}$, satisfy the commutation relations
\begin{equation}
[\phi_{\alpha,R,0},N_{\alpha',R}]=-i\delta_{\alpha,\alpha'},~[\phi_{\alpha,L,0},N_{\alpha',L}]=i\delta_{\alpha,\alpha'}
\label{LRCOMM}.
\end{equation}
For completeness, the mode expansions of the symmetric and anti-symmetric fields, introduced in Eq.~\eqref{SYM_ANTISYM_DEF}, are
\begin{align}
\phi_{\pm}=\phi_{\pm,0}+ \pit_{\pm}\frac{x}{L}+\nonumber \\
\sum_{k\neq0}\sqrt{\frac{1}{2L|k|}}(a_{k,\pm}^{\dagger}e^{ikx}+a^{\phantom{\dagger}}_{k,\pm} e^{-ikx}),
\end{align}
\begin{align}
\theta_{\pm}=\theta_{\pm,0}+ \pi_{\pm}\frac{x}{L}+\nonumber \\
\sum_{k\neq0}\frac{\mathrm{sgn}(k)}{\sqrt{2L|k|}}(a_{k,\pm}^{\dagger}e^{ikx}+a^{\phantom{\dagger}}_{k,\pm} e^{-ikx}),
\end{align}
where we have defined
\begin{subequations}
\begin{gather}
 \phi_{\pm,0}=\left( \frac{(\phi_{1,L,0}+\phi_{1,R,0})\pm(\phi_{2,L,0}+\phi_{2,R,0})}{\sqrt{8\pi}} \right),\\
 \theta_{\pm,0}=\left( \frac{(\theta_{1,L,0}-\theta_{1,R,0})\pm(\theta_{2,L,0}-\theta_{2,R,0})}{\sqrt{8\pi}} \right) , \\
  \pit_{\pm,0} = \sqrt{\pi}\left(\frac{ (N_{1,L}+N_{1,R}) \pm (N_{2,L}+N_{2,R} }{\sqrt{2}} \right),\\
 \pi_{\pm,0} = \sqrt{\pi}\left(\frac{ (N_{1,L}-N_{1,R}) \pm (N_{2,L}-N_{2,R} }{\sqrt{2}} \right), \\
  a_{k,\pm}=\frac{\left( a_{k,1} \pm a_{k,2} \right)}{\sqrt{2}} . 
\end{gather}
\label{DEF}
\end{subequations}
The relevant commutation relations, which can be derived using Eqs.~\eqref{LRCOMM} and \eqref{DEF}, are
\begin{equation}
[\phi_{+,0},\pi_{+,0}]=i,~[\theta_{-,0},\pit_{-,0}]=i,~[\phi_{-,0},\pi_{-,0}]=i.
\end{equation}

We now find the ground state of the total Hamiltonian, Eq.~\eqref{H_TOTAL}. After finding the ground state, we can easily obtain the momentum-space entanglement properties of the model. Following Ref.~\cite{PhysRevB.88.245137}, we expand the cosine terms to quadratic order in field strength. The scalar fields from the quadratic Hamiltonian are then mode expanded and the Hamiltonian is diagonalized via a Bogoliubov transformation. For small values of $J_{\perp}$ and $J^z_{\perp}$, this expansion should be done after the coupling grows under renormalization group flow. As such, our approach is valid when the system size is larger than the correlation length. The analytical predictions obtained from this method have been numerically investigated with exact diagonalization in Ref.~\cite{PhysRevB.88.245137} and Ref.~\cite{2014arXiv1404.7545L}. Excellent agreement was found between the analytical and numerical results. The method developed in Ref.~\cite{PhysRevB.88.245137} was also recently used to study the real-space entanglement spectrum of wire constructions of fractional quantum Hall phases \cite{2014arXiv1411.5369C}.

\subsection{$J_{\perp}\neq0,~J_\perp^z\neq0,~K_-\le\frac{1}{2}$}
We first consider the Hamiltonian in Sec.~\ref{sec:A}. Recall these results also apply for the parameters in Sec.~\ref{sec:C} upon switching $\lambda \pi J_\perp^{xy}/2$ for $J_\perp^z$ in the symmetric channel. We note that due to the locking of the $\phi_+$ and $\theta_-$ fields, the winding modes of these fields must be suppressed, i.e. $\pit_+=\pi_-=0$. We expand the cosine terms as
\begin{equation}
\frac{\pi J_{\perp}a}{(2\pi a)^2} \cos(\sqrt{2\pi}\theta_-)\approx \mathrm{const.}+\frac{u_-m^2_{A,-}K_-}{2}(\theta_--\bar{\theta}_{-,0})^2
\end{equation}
and
\begin{equation}
\frac{J^z_{\perp}a}{(2\pi a)^2} \cos(\sqrt{8\pi}\phi_+)\approx \mathrm{const.}+\frac{u_+m_{A,+}^2}{2K_+} (\phi_+-\bar{\phi}_{+,0})^2
\end{equation}
where $\bar{\phi}_{+,0}$ and $\bar{\theta}_{-,0}$ are the locking positions. Plugging in the mode expansion and writing $H_{\pm}=H^{\mathrm{zero}}_{\pm}+H^{\mathrm{osc}}_{\pm}$, we find for the oscillator portion,
\begin{equation}\label{eq:Hpmosc}
 H^{\mathrm{osc}}_{\pm}=\frac{u_{\pm}}{2} \sum_{k\neq0} 
  \left( a_{k,\pm}^{\dagger},a_{-k,\pm} \right) 
  \begin{pmatrix} A_{k,\pm} & B_{k,\pm} \\ B_{k,\pm} & A_{k,\pm} \end{pmatrix}  
  \begin{pmatrix} a_{k,\pm} \\ a_{-k,\pm}^\dagger \end{pmatrix} 
\end{equation}
with 
\begin{equation}
\begin{split}
 A_{k,+} &= \frac12 \left( K_+ + \frac{1}{K_+} \right) |k|+ \frac{m_{A,+}^2}{2|k|K_+}, \\
 B_{k,+} &= \frac12 \left( -K_+ + \frac{1}{K_+} \right) |k| +\frac{m_{A,+}^2}{2|k|K_+}, \\
A_{k,-} &= \frac12 \left( K_- + \frac{1}{K_-} \right) |k|+\frac{K_-m_{A,-}^2}{2|k|}, \\
 B_{k,-} &= \frac12 \left( -K_- + \frac{1}{K_-} \right) |k| -\frac{K_-m_{A,-}^2}{2|k|}.
\end{split}
\end{equation}
By performing a Bogoliubov transformation
\begin{equation}
 \begin{pmatrix} a_{k,\pm} \\ a_{-k,\pm}^{\dagger} \end{pmatrix}
 =\begin{pmatrix} \cosh \theta_{k,\pm} &  \sinh \theta_{k,\pm} \\ \sinh \theta_{k,\pm} & \cosh \theta_{k,\pm} \\ \end{pmatrix}  
 \begin{pmatrix} b_{k,\pm} \\ b_{-k,\pm}^{\dagger} \end{pmatrix}
\end{equation}
with 
\begin{subequations}
\begin{gather}
 \cosh \left( 2\theta_{k,\pm} \right) = \frac{A_{k,\pm}}{\lambda_{k,\pm}},~~\sinh \left( 2\theta_{k,\pm} \right)=-\frac{B_{k,\pm}}{\lambda_{k,+}},\\
 \lambda_{k,\pm}=\sqrt{A_{k,\pm}^2-B_{k,\pm}^2}=\sqrt{k^2+m_{A,\pm}^2},
\end{gather}
\end{subequations}
the oscillator part of $H_\pm$ is diagonalized as
\begin{equation}
 H_\pm = u_\pm \sum_{k\ne 0} \lambda_{k,\pm} \left( b_{k,\pm}^\dagger b_{k,\pm} + \frac12 \right). 
\end{equation}
We thus have a Klein-Gordon Hamiltonian with a mass gap $u_{\pm}m_{A,\pm}$. Renormalization group calculations to lowest order in $J_{\perp}$ and $J_{\perp}^z$ yields
\begin{equation}
m_{A,+}=\Lambda(\frac{2K_+J_{\perp}^za}{\pi u_+(a\Lambda)^2})^{\frac{1}{2-2K_+}}
\end{equation}
and
\begin{equation}
m_{A,-}=\Lambda(\frac{2 J_{\perp}a}{u_-K_-(a\Lambda)^2})^{\frac{1}{2-\frac{2}{K_{-}}}}
\end{equation}
where $\Lambda$ is a high-energy cut-off. For the details of this calculation, see Ref.~\cite{Giamarchi:book} or ~\cite{Gogolin:book}. The ground state $|0\rangle$ of $H^A_\pm$ is specified by the condition that 
$b_{k,\pm}|0\rangle=0$ for all $k\ne 0$. For the zero mode contribution we find
\begin{equation}
H^{\mathrm{zero}}_+=\frac{u_+}{2}(\pi_+^2\frac{K_+}{L}+\frac{Lm^2_{A,+}}{K_+}(\Delta\phi_{+,0})^2),
\end{equation}
\begin{equation}
H^{\mathrm{zero}}_-=\frac{u_-}{2}(\pit_-^2\frac{1}{LK_-}+Lm_{A,-}^2K_-(\Delta\theta_{-,0})^2).
\end{equation}
where we have defined $\Delta \phi_{\pm,0}=\phi_{\pm,0}-\bar{\phi}_{\pm,0}$ and $\Delta\theta_{\pm,0}=\theta_{\pm,0}-\bar{\theta}_{\pm,0}$.

We are now in position to calculate the oscillator part of the reduced density matrix using methods of free theories originally introduced by Peschel \cite{2003JPhA...36L.205P}. We first calculate the two-point correlation function for right moving particles. This is given by
\begin{equation}
\langle0|a^{\dagger}_{k,\pm}a^{\phantom{\dagger}}_{k,\pm}|0\rangle=\mathrm{sinh}^2\theta_{k,\pm}=\frac{\mathrm{cosh}(2\theta_{k,\pm})-1}{2}.
\end{equation}
Introducing the ansatz,
\begin{equation}
\rho_{A,\pm}^{\mathrm{osc}}=\frac{1}{Z_{e,\pm}^{\mathrm{osc}}}e^{-H_{e,\pm}^{\mathrm{osc}}},~Z_{e,\pm}^{\mathrm{osc}}=\mathrm{Tr}e^{-H_{e,\pm}^{\mathrm{osc}}}.
\end{equation}
with
\begin{equation}
H_{e,\pm}^{\mathrm{osc}}=\sum_{k>0}w_{k,\pm}\left(a^{\dagger}_{k,\pm}a^{\phantom{\dagger}}_{k,\pm}+\frac{1}{2}\right),
\end{equation}
an alternate expression for the two-point correlation function is given by Bose distribution,
\begin{equation}
\mathrm{Tr}(a^{\dagger}_{k,\pm}a^{\phantom{\dagger}}_{k,\pm}e^{-H_{e,\pm}^{\mathrm{osc}}})=\frac{1}{e^{\omega_{k,\pm}}-1}.
\end{equation}
Equating the two expressions for the two-point correlation function, we find
\begin{equation}
w_{k,\pm}=\mathrm{ln}\left(\frac{\mathrm{cosh}(2\theta_{k,\pm})+1}{\mathrm{cosh}(2\theta_{k,\pm})-1}\right).
\label{ES_SLOPE}
\end{equation}
Using Eq.~\eqref{ES_SLOPE}, we find, to lowest order in sub-system momentum,
\begin{align}
w_{A,+,k}=v_{e,A,+}k=\frac{4K_+}{m_{A,+}}k,\nonumber \\
w_{A,-,k}=v_{e,A,-}k=\frac{4}{K_- m_{A,-}}k.
\end{align}
We see that the slope of the entanglement spectrum is non-universal as it depends explicitly on the cut-off. Even the ratio of the  velocities of the symmetric and anti-symmetric depend on the cut-off. The entanglement Hamiltonian for the symmetric (anti-symmetric resp.) oscillator part, is then
\begin{equation}
H_{e,\pm}=v_{e,A,\pm}\left(\sum_{k>0}ka^{\dagger}_{k,\pm}a^{\phantom{\dagger}}_{k,\pm}-\frac{\pi}{12L}\right).
\end{equation}
Here we have used $\zeta$-function regularization $\zeta(-1)=-\frac{1}{12}$ for the infinite constant term.

We now consider the zero-mode part of the entanglement spectrum. The commutation relations for the zero modes allow for the identification of the zero-mode Hamiltonian as a discrete harmonic oscillator. In the limit of large system size, the discreteness of the harmonic oscillator is irrelevant. As such, the ground-state in the $\pi_+,\pit_-$ basis can be approximated as a Gaussian,
\begin{align}
\langle\pi_+,\pit_-|G\rangle=e^{-\frac{K_+(\pi_+)^2}{2m_{A,+}L}-\frac{(\pit_-)^2}{2m_{A,-}LK_-}}.
\end{align}
Using $\pit_+,\pi_-=0$, the reduced density matrix for right moving zero modes is then
\begin{align}
\rho_{A}^{\mathrm{zero}}=\nonumber \\
\sum_{N_{R,1},N_{R,2}}|N_{R,1},N_{R,2}\rangle e^{-\frac{2\pi K_+(N_{R,+})^2}{m_{A,+}LK^2}-\frac{2\pi(N_{R,-})^2}{m_{A,-}LK_-}}\langle N_{R,1},N_{R,2}|.
\end{align}
where $N_{R,\pm}=N_{R,1}\pm N_{R,2}$. The total entanglement Hamiltonian for the zero-mode part is then
\begin{equation}
H_{A,e}^{\mathrm{zero}}=\frac{\pi}{2L}\left(v_{e,A,+}N_{R,+}^2+v_{e,A,-}N_{R,-}^2\right).
\end{equation}
The total entanglement Hamiltonian for right moving particles is then $H_{e}=H_{e,+}+H_{e,-}+H_{e}^{\mathrm{zero}}$. Both $H_{e,-}$ and $H_{e,+}$ are gapless and have a central charge of one, thus $H_{e}$ is gapless and has a central charge of two.

We now calculate the entanglement entropy, $S$, in the large $L$ limit. Following the method first outlined in Ref.~\cite{PhysRevB.88.245137}, we calculate the partition function, which is given by
\begin{equation}
Z_e(\beta)=Z_e^{\mathrm{zero}}(\beta)Z_{e,-}^{\mathrm{osc}}(\beta)Z_{e,+}^{\mathrm{osc}}(\beta)
\end{equation}
We now look at the partition function for the oscillators
\begin{equation}
Z^{\mathrm{osc}}_{e,\pm}(\beta)=e^{\frac{\pi}{24}\tau_{2,\pm}}\prod_{j=1}^{\infty}\left(\frac{1}{1-e^{-2\pi \tau_{2,\pm} j}}\right)=\frac{1}{\eta(i\tau_{2,\pm})}
\end{equation}
where $\eta$ is the Dedekind eta function and $\tau_{2,\pm}=\frac{\beta v_{e,\pm}}{L}$. We are interested in the entanglement entropy as $L$ approaches infinity. As $L\rightarrow\infty$, we have
\begin{equation}
Z^{\mathrm{osc}}_{e,\pm}(\beta)\approx e^{\frac{LT\pi}{v_{e,\pm} 24}}\sqrt{\frac{v_{e,\pm}}{LT}},
\end{equation} 
The partition function for the symmetric zero-mode channel is
\begin{equation}
Z_{e}^{\mathrm{zero}}=\sum_{N_{1,R},N_{2,R}}e^{-\frac{\beta\pi}{2LK}( \bold{N}_R^T\Omega\bold{N}_R)}=\sum_{N_{1,R},N_{2,R}}e^{-\pi \tau ( \bold{N}_R^T\Omega\bold{N}_R)}
\end{equation}
where 
\begin{align}
 \Omega=
  \begin{pmatrix}  v_{e,A,+}+v_{e,A,-}&v_{e,A,+}-v_{e,A,-} \\ v_{e,A,+}-v_{e,A,-}  & v_{e,A,+}+v_{e,A,-}\end{pmatrix},
\end{align}
$\bold{N}_R=\begin{pmatrix} N_{1,R},~N_{2,R}\end{pmatrix}$ and $\tau=\frac{\beta }{2L}$. The partition function for zero modes is the Riemann theta function. Using the modular properties of the Riemann theta function, in the large $L$ limit we find
\begin{equation}
Z_{e,A}^{\mathrm{zero}}=\frac{\theta(0|i\tau^{-1}\Omega^{-1})}{\sqrt{\mathrm{det}(\tau\Omega)}}\approx \left(\frac{v_{e,A,+}v_{e,A,-}\beta^2}{L^2K^2}\right)^{-\frac{1}{2}}
\end{equation}
The total partition function is then
\begin{equation}
Z_{e,A}=e^{\frac{LT\pi}{24}(\frac{1}{v_{e,+}}+\frac{1}{v_{e,-}})}
\end{equation}
The momentum-space entanglement entropy for this partition is then
\begin{align}
S=\frac{\partial (T\mathrm{ln}Z_{e,A})}{\partial T}\large{|}_{T=1}=L\frac{\pi}{12}\left(\frac{v_{e,A,+}+v_{e,A,-}}{v_{e,A,+}v_{e,A,-}}\right).
\end{align}
We thus see we have a volume law. We note that the coefficient of the volume term is non-universal as it depends on the cut-off.

\subsection{$J_\perp^z\neq0,~K_->\frac{1}{2}$}
We now consider the Hamiltonian in Sec.~\ref{sec:B}. If the anti-symmetric channel is gapped, we expand that cosine term in the anti-symmetric channel as
\begin{equation}
\frac{J^z_{\perp}a}{(2\pi a)^2} \cos(\sqrt{8\pi}\phi_-)\approx \mathrm{const.}+\frac{u_-m_{B,-}^2}{2K_-} (\phi_--\bar{\phi}_{-,0})^2
\end{equation}
Due to the locking of the $\phi_+$ and $\phi_-$ fields, we have $\pit_-=\pit_+=0$. After plugging in the mode expansions, we find for the oscillator contribution
\begin{equation}\label{eq:Hpmosc}
 H^{\mathrm{osc}}_{\pm}=\frac{u_{\pm}}{2} \sum_{k\neq0} 
  \left( a_{k,\pm}^{\dagger},a_{-k,\pm} \right) 
  \begin{pmatrix} A_{k,\pm} & B_{k,\pm} \\ B_{k,\pm} & A_{k,\pm} \end{pmatrix}  
  \begin{pmatrix} a_{k,\pm} \\ a_{-k,\pm}^\dagger \end{pmatrix} 
\end{equation}
with 
\begin{equation}
\begin{split}
 A_{k,+} &= \frac12 \left( K_+ + \frac{1}{K_+} \right) |k|+ \frac{m_{B,+}^2}{2|k|K_+}, \\
 B_{k,+} &= \frac12 \left( -K_++ \frac{1}{K_+} \right) |k| +\frac{m_{B,+}^2}{2|k|K_+}, \\
A_{k,-} &= \frac12 \left( K_-+ \frac{1}{K_-} \right) |k|+\frac{m_{B,-}^2}{2|k|K_-}, \\
 B_{k,-} &= \frac12 \left( -K_- + \frac{1}{K_-} \right) |k| -\frac{m_{B,-}^2}{2|k|K_-}.
\end{split}
\end{equation}
For this case, we again have a sine-Gordon model with a mass gap. Renormalization group calculations to lowest order in $J_{\perp}$ and $J_{\perp}^z$ yield
\begin{equation}
m_{B,+}=\Lambda(\frac{2K_+J_{\perp}^za}{\pi u_+(a\Lambda)^2})^{\frac{1}{2-2K_+}}
\end{equation}
and
\begin{equation}
m_{B,-}=\Lambda(\frac{2K_- J_{\perp}^za}{\pi u_-(a\Lambda)^2})^{\frac{1}{2-2K_-}}
\end{equation}
We find the zero-mode contribution to be
\begin{equation}
H^{\mathrm{zero}}_+=\frac{u_+}{2}(\pi_+^2\frac{K_+}{L}+\frac{Lm^2_{B,+}}{ K_+}(\Delta\phi_{+,0})^2),
\end{equation}
\begin{equation}
H^{\mathrm{zero}}_-=\frac{u_-}{2}(\pi_-^2\frac{K_-}{L}+\frac{Lm^2_{B,-}}{K_-}(\Delta\phi_{-,0})^2).
\end{equation}

Following similar steps as the previous section, we find the entanglement Hamiltonian for the oscillators to be
\begin{equation}
H_{e,B,\pm}^{\mathrm{osc}}=v_{e,B,\pm}\left(\sum_{k>0}ka^{\dagger}_{k,\pm}a^{\phantom{\dagger}}_{k,\pm}-\frac{\pi}{12L}\right).
\end{equation}
where
\begin{align}
v_{e,B,+}=\frac{4K_+}{ m_{B,+}},~v_{e,B,-}=\frac{4K_-}{m_{B,-}}.
\end{align}
Using $\pit_{+},\pit_-=0$, we find the zero-mode contribution of the entanglement Hamiltonian to be
\begin{equation}
H_{B,e}^{\mathrm{zero}}=\frac{\pi}{2L}\left(v_{e,B,+}N_{R,+}^2+v_{e,B,-}N_{R,-}^2\right).
\end{equation}
Evaluating the partition function in the same manner as the previous section, we find the entanglement entropy to be
\begin{equation}
S=L\frac{\pi}{12}\left(\frac{v_{e,B,+}+v_{e,B,-}}{v_{e,B,+}v_{e,B,-}}\right).
\end{equation}

We can also treat the small gapless regime of the anti-symmetric mode at $J_{\perp}^{xy}=0$ and $K>1-\frac{J_{\perp}^z}{\pi u_-}$. One can show in this case (by neglecting the irrelevant cosine term) the entanglement spectrum for the anti-symmetric mode is flat and $w_{k,-}^B$ is given by
\begin{equation}
w_{k,-}^B=\mathrm{ln}\left(\frac{K_-+(K_-)^{-1}+2}{K_-+(K_-)^{-1}-2}\right).
\end{equation}
The symmetric part of the entanglement Hamiltonian remains gapless, thus we have an entanglement Hamiltonian with one gapless mode and one dispersion-less  mode.
\section{Conclusion}
We have studied the momentum-space entanglement entropy and spectrum of anisotropic Heisenberg spin-half ladders using field theoretical methods. We found that the entanglement entropy between left and right movers is linear in system size and obeys a volume law. When the system is gapped, the momentum-space entanglement Hamiltonian was found to be gapless and described by a conformal field theory with a central charge of two. If the system is gapless, the entanglement Hamiltonian was found to be described by one dispersion-less mode and one gapless mode with a linear spectrum. Our work can easily be generalized to include exchange between the spins across diagonals of the plaquettes and the following four-spin term $(S_{1,i}S_{1,i+1}S_{2,i}S_{2,i+1})$, which can arise due to phonons. We note exact diagonalization suffers significantly from finite size effects for weak coupling, as the typical correlation length is larger than system sizes currently available. As such, it would be interesting to numerically investigate the entanglement spectrum via quantum monte carlo methods, which have recently been applied to entanglement spectra studies \cite{PhysRevB.89.195147,2014PhRvB..90l5105L,PhysRevB.90.081116}, or the momentum-space density matrix renormalization group. Finally, it would also be interesting to generalize our work to study other gapless phases of spin ladders, including the vector chiral phase \cite{PhysRevLett.81.910}, and inequivalent chains.

\acknowledgments

We thank P. Laurell, R. Thomale and R. Santos for useful discussions. We also thank S. Furukawa and G. Fiete for a critical reading of the manuscript and collaboration on related work. R.L. was supported by National Science Foundation Graduate Research Fellowship award number 2012115499 and NSF Grant No. DMR- 0955778.

\bibliography{EScTLL.bib}

\end{document}